\documentstyle[epsfig,aasms4]{article}

 1

\def\Msun{M_\odot}
\def\Lsun{L_\odot}

\def\lta{\lower.5ex\hbox{\ltsima}}
\def\gta{\lower.5ex\hbox{\gtsima}}
\def\ltsima{$\; \buildrel < \over \sim \;$}
\def\lsim{\lower.5ex\hbox{\ltsima}}
\def\gtsima{$\; \buildrel > \over \sim \;$}
\def\gsim{\lower.5ex\hbox{\gtsima}}

\def\upm{{\stackrel{\raise.5ex\hbox{$m$}}{\lower.2ex\hbox{.}}}}
\def\upd{{\stackrel{\raise.5ex\hbox{$d$}}{\lower.2ex\hbox{.}}}}

\def\gs{\mathrel{\raise1.16pt\hbox{$>$}\kern-7.0pt
\lower3.06pt\hbox{{$\scriptstyle \sim$}}}}
\def\ls{\mathrel{\raise1.16pt\hbox{$<$}\kern-7.0pt
\lower3.06pt\hbox{{$\scriptstyle \sim$}}}}
\def\gtsima{$\; \buildrel > \over \sim \;$}
\def\ltsima{$\; \buildrel < \over \sim \;$}
\def\prosima{$\; \buildrel \propto \over \sim \;$}
\def\gsim{\lower.5ex\hbox{\gtsima}}
\def\lsim{\lower.5ex\hbox{\ltsima}}
\def\simgt{\lower.5ex\hbox{\gtsima}}
\def\simlt{\lower.5ex\hbox{\ltsima}}
\def\simpr{\lower.5ex\hbox{\prosima}}

\def\pp{\\parshape 2 0truecm 17truecm 2truecm 15truecm}
\def\rf#1;#2;#3;#4 {\par\pp#1, #2, #3, #4. \par}

\def\pr{\ref@jnl{Phys.Rev}}

\def\href#1;#2 {{\bf #1} : {\em #2}}

\def\beq#1{\begin{equation}\label{#1}}
\def\eeq{\end{equation}}
\def\beqa#1{\begin{eqnarray}\label{#1}}
\def\eeqa{\end{eqnarray}}



\begin{document}
\thispagestyle{empty}
\title{\huge{\bf Binary\ Masses\ as\ a\ Test\ for
Pre--Main-Sequence\ Tracks\ }}
 
\bigskip
\bigskip
\author {\large Francesco Palla\altaffilmark{1} and 
Steven W. Stahler\altaffilmark{2} }

\altaffiltext{1}{Osservatorio Astrofisico di Arcetri, L.go E. Fermi 5, 
50125 Florence, Italy - palla@arcetri.astro.it}
\altaffiltext{2}{Berkeley Astronomy Department, University of California,
Berkeley, Ca 94720, USA - Sstahler@astro.berkeley.edu}

\slugcomment{To appear in {\it The Astrophysical Journal}, May 20, 2001}
%\received{March 15, 1999}
%\accepted{---------------}

\begin{abstract}

Observations of binaries have traditionally provided the means for
ascertaining stellar masses. Here, we use the published data on 8
pre--main-sequence pairs to gauge the accuracy of our own, recently
calculated, evolutionary tracks (Palla \& Stahler 1999). We consider
both eclipsing, double-lined spectroscopic binaries, which provide the
mass of each star separately, and non-eclipsing, double-lined systems,
which yield only the ratio. We also analyze the visual, quadruple
system GG Tau, for which the sum of the two component masses follows from
observations of the circumbinary disk.

In almost all cases, our theoretically derived masses or mass ratios are
in good agreement with the empirical values. For two binaries
(NTTS 162814$-$2427 and P1540), the observational results are still too
uncertain for a proper comparison. We also find that the derived contraction
ages within each pre--main-sequence pair are nearly equal. This result
extends earlier findings regarding visual pairs, and indicates that the
components of all binaries form in proximity, perhaps within the same dense
cloud core. Finally, our study reveals that the Trapezium star BM Ori is
very young, since both the star itself and its companion have contraction
ages less than $10^5$ years.

\end{abstract}

%\twocolumn 

\section{Introduction}

Pre-main-sequence stars derive their luminosity from the gravitational
contraction they experience prior to hydrogen ignition. Within the HR
diagram, they descend evolutionary tracks that start from the birthline and
end at the zero-age main sequence. Although these objects emit vigorous winds
and undergo disk accretion, especially at earlier times, the net change in
mass during the contraction phase appears to be relatively small (e.g.,
Hartmann et al. 1998). Thus, it is still appropriate to consider the ensemble
of tracks parametrized by the stellar mass.

The masses and ages derived by placing stars on the tracks are fundamental
data, not only for characterizing individual objects, but for assessing the
mass distribution and history of formation activity within clusters and
associations. Following the initial calculations of the 1960s, Cohen \& Kuhi
(1979) compiled a set of tracks that were used for many years. Swenson et
al.  (1994) and D'Antona \& Mazzitelli (1994) then offered new sets that
utilized different phenomenological theories of convection. These later
results not only differed substantially from each other, but also led to
problematic distributions of stellar masses and ages (e.g. Green \& Meyer
1995; Hillenbrand 1997). A number of authors have reinvestigated the problem,
in part to resolve these differences (Mart\'{\i}n \& Claret 1996; Chabrier \&
Baraffe 1997; Baraffe et al. 1998; D'Antona \& Mazzitelli 1998; Siess et al.
1999). Our own recent calculation (Palla \& Stahler 1999) utilized standard
mixing-length theory for convection, along with modern determinations of the
equation of state and low-temperature opacities.

How are we to gauge the accuracy of pre-main-sequence tracks? One test is the
position of the birthline, i.e., the upper envelope to the distribution of
pre-main-sequence stars. Stahler (1983) first predicted this locus
theoretically by combining a protostellar mass-radius relationship with the
Cohen \& Kuhi tracks. Palla \& Stahler (1990) then extended the curve into
the regime of intermediate-mass objects. Thus far, the predicted birthline
does seem to match the observed envelopes in both T associations and groups
containing Herbig Ae/Be stars (e.g., Chen et al. 1997; van den Ancker et al.
1998; Luhman 1999). This result attests to the essential validity of both
protostar theory (leading to the mass-radius relation) and the
pre-main-sequence tracks themselves.

For a more precise check, one should compare the stellar masses obtained
using the tracks with those that can be derived empirically. On the
observational side, there has been some attempt to derive masses from
spectroscopically determined surface gravities (MacNamara 1975; 1976), but
the precision of the latter is rather poor. Bonnell et al. (1998) suggested
using the infall speed of circumstellar matter, as gauged from the redshifted
absorption of selected emission lines. The presumption here is that the
detected motion occurs along magnetic flux tubes linking the inner edge of a
disk to the stellar surface.  Quantitatively, however, one faces not only the
problem of geometric projection along field lines of unknown topology, but
also the complication that the physical velocities may not be truly
free-fall. The most reliable technique is to use {\it rotational} speeds.
Here, one may employ the radial velocities of spectroscopic binaries (e.g.,
Lee 1992), or else molecular-line observations of disks around binaries
(White et al. 1998) or single stars (Simon et al. 2000).

In this paper, we employ both spectroscopic binaries and circumbinary disks
to gauge the accuracy of our own pre-main-sequence tracks.  We begin, in \S
II below, by verifying the endpoints of these tracks, through observations of
main-sequence systems.  We then consider, in \S III, a number of younger
pairs. We focus initially on eclipsing, spectroscopic binaries, for which
individual masses can be obtained with high accuracy.  We analyze several
systems that contain at least one pre-main-sequence member.  Mass ratios, as
opposed to individual values, are available for double-lined spectroscopic
binaries, from the relative amplitudes of the individual velocity curves.
This ratio may be compared with pre-main-sequence theory, provided one can
also assess the luminosities and effective temperatures of the binary
components. We finally consider the special case of GG Tau, where combined
stellar masses can be derived from the rotational velocity of the
circumbinary disk.

Our conclusion, as presented in \S IV, is that the pre-main-sequence tracks
in question are indeed reliable, in the sense that they generally predict
masses or mass ratios in excellent agreement with empirical values. Moreover,
we find that the derived ages within a pair are usually very similar. Close
binaries thus appear to be coeval, a tendency already adduced for wider
systems (Hartigan et al. 1994; Brandner \& Zinnecker 1997). We find two
anomalous systems, NTTS~162814$-$2427 and P1540, for which both the mass
ratios and ages seem discrepant. The fault, we suggest, lies in the
determination of the surface temperatures and luminosities, a conclusion
reinforced by other observations.

\section{Main-Sequence Binaries}

Binary stars have traditionally been the main source of accurate stellar
masses and radii, and are employed to calibrate the mass-dependence of the
main sequence (Andersen 1991). Most useful in this regard are the detached,
double-lined, eclipsing systems. Here, one exploits both the radial velocity
and light curves to obtain not only the masses of the individual components,
but also their luminosities and effective temperatures.

Figure 1 shows a portion of our pre--main-sequence tracks in the HR diagram,
with masses from 0.1 to 1.2 $\Msun$. Toward the upper end we have placed the
components of two eclipsing systems with primaries of roughly solar mass, EW
Ori and HS Aur. It is evident that the measured luminosities and effective
temperatures indeed place all four stars close to the end of our tracks. From
Table 1 of Andersen (1991), the masses of the EW Ori components are 1.19 and
1.16 $\Msun$.  For comparison, the interpolation from our tracks yields
values of 1.14 and 1.11 $\Msun$, respectively. For HS Aur, the measured
masses are 0.90 and 0.88 $\Msun$, while our tracks give corresponding values
of 0.94 and 0.90 $\Msun$, respectively.

The number of known systems with lower-mass components diminishes rapidly.
Until recently, there were only two observed eclipsing binaries with subsolar
primaries, YY Gem and CM Dra.  The latter has a metallicity from a third to
half of solar (Leggett et al. 2000), and thus cannot be studied using our
tracks. (For the effect of reduced metallicity on evolutionary tracks of
low-mass stars, see, e.g., D'Antona \& Mazzitelli 1998, Siess et al. 2000).
Delfosse et al. (1999) have discovered a third system, GJ 2609, a quadruple
whose more massive component is itself an eclipsing binary. No estimates for
effective temperature are yet available.

The measured masses of YY Gem are 0.60 and 0.59 M$_\odot$ (Leung \& Schneider
1978; S\'egransan et al. 2000). We display the positions of both components
in Figure 1. Note the large uncertainties in effective temperature, a well
recognized problem at these masses (Allard et al. 1997).  Our adopted T$_{\rm
eff}$-values for YY Gem (3806 and 3742 K) come from Chabrier \& Baraffe
(1995). These nominal values, together with the measured luminosities, again
place the stars close to, but not precisely at the ends of our tracks. Our
inferred masses, 0.57 and 0.49 $\Msun$ for YY Gem are also discrepant than
for solar-mass systems, although the error is still under 20\% percent. Such
declining accuracy at the lowest masses is expected, as our models lack the
refined equation of state and atmosphere calculations needed for accurate
treatment in this regime (eg, Chabrier \& Baraffe 1995).

\section{Younger Systems}

\subsection{Rs~Cha}

This eclipsing, double-lined spectroscopic binary belongs to the recently
discovered open cluster $\eta$ Cha. The group itself contains about 50
members, and lies at a distance of just 97~pc (Mamajek et al. 1999). Strong
X-ray emission from these stars identifies them as pre-main-sequence (Mamajek
et al. 2000). For the RS~Cha system, which has a period of 1.7 days, the
spectroscopic and photometric analysis of Andersen (1991) yielded primary and
secondary masses of 1.86~$\Msun$ and 1.82~$\Msun$, respectively, with an
associated mass ratio of \hbox{$M_1/M_2\equiv q = 1.02$}.  Although this
value is very close to unity, we have included the system in our sample
because its eclipses allow accurate determination of individual masses.

Jordi et al. (1997) obtained effective temperatures of 7810~K and 7295~K for
the two components. Mamajek et al. (2000) used these results, along with the
stellar radii determined by Clausen \& Nordstr\"om (1980), to find
corresponding luminosities of 15.2~$\Lsun$ and 13.9~$\Lsun$. We show the two
stars within the HR diagram in Figure 2a. Our derived masses of 1.88~$\Msun$
and 1.80~$\Msun$ agree well with the empirical ones. Our theoretical
$q$-value is 1.04. Finally, the derived ages for the primary and secondary
are $5.0\times 10^6$~yr and $4.3\times 10^6$~yr, respectively.

As an aside, we note that the stellar parameters of the two components place
them near the regime of pulsational instability determined by Marconi \&
Palla (1998). Fig.~4 shows that the secondary falls securely within the
theoretical instability strip, indicated by the two parallel lines. This
object should therefore be pulsating in the fundamental or first overtone.
Any pulsation of the primary, however, would be of higher order. Andersen
(1975) observed, in fact, a periodic variation (of the $\delta$~Scuti type)
in the composite light curve. Further comparison between theory and
observation should prove instructive.

\subsection{EK~Cep}

This system lies at a distance of 150~pc, and does not appear to be
associated with other young stars (Popper 1987). Ebbighausen (1966) showed it
to be an eclipsing, double-lined binary of low eccentricity ($e\,=\,0.09$)
and relatively short period (4.4~days). From the radial velocity curves, the
mass ratio is 1.81. Tomkin (1983) derived an orbital inclination $i$ of
$89^\circ$, and thereby found primary and secondary masses of $2.03\,\Msun$
and $1.12\,\Msun$, respectively.

The primary is a ZAMS star of spectral type A1 and effective temperature
9000~K (Hill \& Ebbighausen 1984; Popper 1987). Its luminosity, as determined
photometrically, is $15~\Lsun$. Mart\'{\i}n \& Rebolo (1993) detected lithium
absorption in the secondary, establishing its pre-main-sequence nature. The
effective temperature for this component is 5700~K and its luminosity
$1.5\,\Lsun$ (Popper 1987).

After locating both stars in the HR diagram (Figure 2b), we confirm that the
primary lies on the ZAMS, and find a mass of $2.0\,\Msun$. For the secondary,
we derive $M_\ast\,=\,1.1\,\Msun$, so that our $q$-value is 1.7. The
isochrone passing through the secondary's position gives it an age of
$2\times 10^7$~yr. This figure exceeds the contraction time of $8\times
10^6$~yr for a $2\,\Msun$ star to reach the ZAMS. Thus, the stars could have
formed at the same time, $2\times 10^7$~yr in the past.

\subsection{TY~CrA} 

The Herbig Be star TY~CrA is the brightest member of a young, embedded cluster
in the Corona Australis dark cloud complex, at a distance of 130~pc (Marraco
\& Rydgren 1981). Photometric and spectroscopic observations reveal an
eclipsing, double-lined spectroscopic system (Kardopolov et al. 1981; Corporon
et al. 1994; Casey et al. 1995). Analysis of the radial velocity curves gives
a period of 2.89 days, and shows the orbit to be nearly circular ($e~=~0.02$).
From the relative amplitudes of the velocities, the mass ratio $q$ is 1.93.
The system also contains a third component, at a distance of about 1~AU from 
the close binary (Casey et al. 1995; Corporon et al. 1996).

Casey et al. (1998) have conducted a detailed investigation of the stellar
properties of the binary components. (For earlier work, see Beust et al.
1997.) They assign the primary an effective temperature of $1.2\times
10^4$~K, consistent with the estimated spectral type of B8 (Lagrange et al.
1993).  For the secondary, they find $T_{\rm eff}\,=\,4900$~K. Gauging the
luminosities is more difficult, both because of the third star and the
presence of reflection nebulosity. Casey et al. find primary and secondary
luminosities of $67~\Lsun$ and $2.4~\Lsun$, respectively. Both the
temperature and luminosity estimates utilize an estimated $A_V$ of 3.1~mag.

Figure 2c shows the location of both components in the HR diagram. The primary
evidently falls close to the ZAMS, and has a theoretically derived mass of
$2.9~\Msun$. For the secondary, we infer a mass of $1.6~\Msun$. These figures
may be compared with the masses of $3.16~\Msun$ and $1.64~\Msun$ obtained by
Casey et al. (1998) from the known orbital inclination of $i\,=\,83^\circ$.
Our theoretical mass ratio, $q\,=\,1.8$, differs by less than 10~percent from
the empirical value.

Since the primary may already have settled onto the main sequence, a
contraction age may not be appropriate. A star of $3\,\Msun$ reaches the ZAMS
in $2\times 10^6$~yr, according to our tracks. As seen in Fig.~2c, the
corresponding isochrone passes close to the secondary, indicating near
coevality. However, the nominal contraction age that best fits the
secondary's position in the diagram is $3.9\times 10^6$~yr. The location of
the primary in the diagram is, of course, also consistent with this latter
age.

\subsection{BM~Ori} 

The faintest member of the Trapezium is $\theta^1$~Ori~B, also known as
BM~Ori. It has long been recognized as the primary of an eclipsing binary,
with a period of 6.5 days and a circular orbit (Hartvig 1920; Popper \&
Plavec 1976). More recently, near-infared studies have demonstrated that this
spectroscopic binary is actually part of a hierarchical, pentuple system.
There is a wide binary whose components lie $0.^{\prime\prime}94$ and
$1.^{\prime\prime}02$ from the brightest star (Petr et al. 1998; Weigelt et
al.  1999), and a single star at a displacement of $0.^{\prime\prime}9$
(Simon et al. 1999).

Within the spectroscopic binary, the primary appears to be a normal B3-type
star lying close to the ZAMS (Popper 1980; Hillenbrand 1997). The secondary,
however, shows a large infrared excess, and its spectral classification has
been uncertain for a long time (e.g., Popper \& Plavec 1976). Vitrichenko \&
Plachinda (2000) employed CCD spectrophotometry during the primary minimum and
recovered a G2-type spectrum for the secondary. Antokhina et al. (1989) had
earlier obtained a luminosity of 31~$\Lsun$ for this component.

Figure 2d displays the two stars within the HR diagram. The primary, as
expected, lies close to the ZAMS, and has an inferred mass of $6.2~\Msun$.
The secondary lies very close to the birthline, shown by the dashed curve. We
derive a secondary mass of $2.7~\Msun$ and a $q$-value of 2.3, the highest of
any system we have analyzed. Since the primary also lies close to the
birthline, the binary components are coeval, with nominal ages less than
about $10^5$~yr.

Obtaining the masses from the light curve is not straightforward, because of
the peculiar nature of the system. Even during primary minimum, the B star
continues to be visible (Hall \& Garrison 1969; Doremus 1970). The current
interpretation is that the eclipse in this phase is produced by a disk
surrounding the secondary (Vitricenko 1998). The disk is also presumed
responsible for the infrared excess. In any case, a more conventional
light-curve analysis yields masses of 6.3 and $2.5~\Msun$, respectively, for
the primary and secondary, and a $q$-value of 2.5. A refined study,
accounting for the disk geometry, could alter these numbers.

\subsection{V773~Tau} 

We next consider non-eclipsing, spectroscopic binaries. Our first
system is located in the B209 (L1495W) dark cloud, at an astrometric 
distance of 148~pc (Lestrade et al. 1999). The primary is a weak-lined T Tauri
star that exhibits strong X-ray activity (Feigelson et al. 1994). It is also 
the most luminous radio source in Taurus-Auriga (O'Neal et al. 1990), with 
giant loops of nonthermal emission (Feigelson et al. 1994).

Near-infrared imaging by Leinert et al. (1993) and Ghez et al. (1993) revealed
a wide companion. These two groups found the separation to be 19~AU and 29~AU,
respectively, while Ghez et al. (1997) lowered the figure to 12~AU. In any
case, Mart\'{\i}n et al. (1994) and Welty (1995) demonstrated that the
system is actually a triple, since the brightest source is a spectroscopic
binary. The analysis of the velocity curves by Welty shows the orbit to be
eccentric ($e = 0.267$), and yields a mass ratio of
\hbox{$M_1/M_2\equiv q = 1.32$}. Welty then fit the composite optical spectrum
with K2 and K5 photospheres, weighted to match the observed flux ratio at 
0.65~$\mu$m. He finally determined luminosities for the two components by 
requiring that their spectral energy distributions sum to the empirical one, 
within the optical and near-infrared regime. His derived luminosities, scaled
to the recent, astrometric distance, are $L_1 = 3.01\,\Lsun$ and 
$L_2 = 1.90\,\Lsun$.

These values for luminosity and temperature, in combination with our
pre-main-sequence tracks, yield primary and secondary masses of $1.73\,\Msun$ 
and $1.26\,\Msun$, respectively. Their ratio is 1.37, in good agreement with 
the dynamical value. In contrast, Welty used the tracks of D'Antona \& 
Mazzitelli (1994) and found a theoretical ratio of 1.78. 

Welty's derivation of the luminosities neglected any contribution to the
optical or near-infrared fluxes from the wide companion. This assumption,
however, may not be justified. Ghez et al. (1997) studied the companion
photometrically, and assigned it a spectral type of K3 and a luminosity of 
$1.33~\Lsun$. They then subtracted its spectral energy distribution from the
total, assuming the remainder stems from the spectroscopic binary, at least
for optical and near-infrared wavelengths. Adopting Welty's spectral types
and near-infrared flux ratio for the binary components, they derived new
luminosities of $L_1 = 2.31\,\Lsun$ and $L_2 = 1.39\,\Lsun$. 

Figure~3a shows the two components of the spectroscopic binary in the HR
diagram, using the stellar parameters of Ghez et al (1997). Also shown are
portions of our pre-main-sequence tracks and isochrones. Our derived masses
are now $1.53~\Msun$ and $1.19\,\Msun$, yielding a mass ratio of 1.29. The
ages are $4.1\times 10^6$~yr and $2.8\times 10^6$~yr for the primary and
secondary, respectively.

\subsection{NTTS~162814$-$2427}

The primary of this system was discovered as an X-ray source in the
Scorpius-Ophiuchus region (Montmerle et al. 1983), and later classified as a
naked T Tauri star (Walter et al. 1988). Mathieu et al. (1989) demonstrated
that the star is actually a double-lined spectroscopic binary, with a period
of 36.0 days and a mass ratio $q$ of 1.09. It's measured eccentricity of 0.48
is one of the highest values known among pre-main-sequence binaries.

Although the ultraviolet and infrared excesses in the composite spectrum are
relatively small, they foiled the attempt by Mathieu et al. to deconvolve the
component stars. This problem was alleviated by Lee et al. (1994), who used
high-dispersion spectroscopy both to gauge the stars' relative brightness and
to assign a plausible range of spectral types (see also Lee 1992). For each
pair of spectral types, Lee et al. reconstructed the stellar luminosities by
applying a bolometric correction to the dereddened $I$-band fluxes. Here,
they utilized a visual extinction $A_V$ of 1.9.

The closed areas in Figure~3b represent the four pairs of luminosity and
effective temperature offered by Lee et al. (1994). The effective
temperatures derived by these authors correspond to spectral types of K4 and
K5, after using the conversion scale of Cohen \& Kuhi (1979). Comparison with
our pre-main-sequence tracks shows that both masses are slightly above
$1\,\Msun$, and that the theoretical $q$-values range from 1.04 to 1.18.
These figures neatly bracket the empirical one of 1.09. Fig.~3b also
includes three selected isochrones. Although none of these intersects both
rectangular areas, the data are consistent with a common age near $5\times
10^6$~yr.

\subsection{P1540}

This double-lined, spectroscopic binary is located $10^\prime$ west of the
Trapezium, just outside the Orion Nebula cluster (Hillenbrand 1997).
Marschall \& Mathieu (1988) analyzed the radial velocities, and found the
system to have a relatively long period (34~days) and high eccentricity ($e =
0.12$). The derived mass ratio, $q = 1.32$, matches that of V773~Tau. As in
that case, the components are both weak-lined T~Tauri stars, and exhibit
enhanced X-ray emission. The system is notable for its high space motion.
While the center-of-mass radial velocity of 26~km~s$^{-1}$ is consistent with
membership in the Orion~Id association, the measured proper motion is much
too great (Jones \& Walker 1988). Marschall \& Mathieu conclude that the
system is a runaway, possibly ejected through dynamical interaction within
the Trapezium region.

Observational determination of the component luminosities and effective
temperatures is still incomplete. Lee et al. (1994) have suggested four pairs
of values, in which the spectral types lie in the range from K4 to K5 and the
luminosities are between 6.0 and 11.5~$\Lsun$. The analysis was similar to
that of NTTS~162814$-$2427. That is, a high-resolution spectrum was compared to
comnbinations of standard templates in order to assess both the individual
spectral types and the flux ratio, here in the $I$-band. After using
photometric observations of color excess to correct for extinction, Lee et
al. applied a main-sequence bolometric correction to obtain the stellar
luminosities.

The quadrangles in Figure~3c show the regions demarcated by these four
solution pairs in the HR diagram. It is evident that the system is extremely
young, with the stellar parameters placing both components close to the
birthline.  The derived mass ratios range from 1.0 to 1.37, and thus are
ostensibly compatible with the observational value. However, it is evident
from the figure that none of the pairs are even roughly coeval. This
peculiarity was emphasized by Marschall \& Mathieu (1988), who based their
analysis on the Cohen \& Kuhi (1979) tracks.

Both components of P1540 exhibit strong lithium absorption, another
indication of youth. Lee et al. (1994) employed stellar atmosphere models to
determine the actual lithium abundances of five pre-main-sequence binaries,
including this one. In general, the derived abundance in a star is sensitive
to both the effective temperature and surface gravity, and hence the stellar
luminosity.  For all four luminosity-temperature pairs, P1540 had lithium
abundances for both components higher by a factor of 3--7 than the canonical
interstellar value. Moreover, this was the only system to exhibit such an
anomaly. Lee et al.  suggested that the result stemmed from incorrect stellar
parameters. In particular, a lowering of both spectral types by several
subclasses appears to be compatible with the spectroscopic data (Mart\'{\i}n
2000, personal communication).  Lowering the spectral types would also
decrease the masses, in concordance with the relatively high lithium
abundances.  A detailed reexamination of this interesting system is clearly
warranted.

\subsection{GG~Tau}

The GG~Tau system is a hierarchical quadruple, consisting of two binary stars
(Leinert et al. 1993). The tighter pair, designated GG~Tau~A, has a projected
separation of $0.^{\prime\prime}25$, or 35~AU at the Taurus distance. This
system is $10.^{\prime\prime}1$ north of a wider binary (GG~Tau~B), with a
separation of $1.^{\prime\prime}48$ (207~AU). Thus, neither pair is a
spectroscopic binary. However, the system is of interest for testing
pre-main-sequence models because of the presence of a large, low-mass disk
around GG~Tau~A. This structure has been imaged in both millimeter lines and
continuum, and in near infrared scattered light (Dutrey et al. 1994; Roddier
et al. 1996). Following the initial study by Dutrey et al. (1994), Guilloteau
et al. (1999) have fit the disk's velocity profile to a Keplerian rotation
low, thus finding the {\it total} mass of GG~Tau~A to be 1.28~$\Msun$.

White et al. (1999) have obtained optical spectra for all members of the
quadruple. They confirm that the components are T~Tauri stars or brown
dwarfs, with spectral types ranging from K7 to GG~Tau~Aa to M7 for
GG~Tau~Bb.  They further derive effective temperatures for all the stars, and
estimate their luminosities by applying a bolometric correction to the
$I$-band fluxes. Figure~3d places the components in the HR diagram. The two
most luminous stars belong to GG~Tau~A, and our derived masses are 0.78
$\Msun$ for the primary and 0.54 $\Msun$ for the secondary.  The total is
$1.32~\Msun$, in close agreement with the dynamical value.  Both
components of GG~Tau~B lie below our tracks, and we can only place upper
limits of $0.1~\Msun$ on their masses. Note finally that all four stars
appear to be approximately coeval. Our derived ages for GG~Tau~Aa and
GG~Tau~Ab are, respectively, 2.3 and 1.5$\times 10^6$~yr.

\section{Discussion}

Table 1 conveniently summarizes essential properties of all the systems
covered in this paper. We first list empirical values for the effective
temperature and luminosity of each component, including uncertainties
given in the literature. We next give the observationally determined
mass ($M_{\rm dyn}$) for all eclipsing systems, as well as the total
value for GG~Tau~A. In the fifth column, we tabulate our own masses
($M_{\rm PS}$), as obtained by comparison with the pre-main-sequence
tracks. The next two columns list the mass ratios, both the observational
ones ($(M_A/M_B)_{\rm dyn}$) and our derived values ($(M_A/M_B)_{\rm PS}$).
The final column presents our derived ages, again including uncertainties,
for each component that appears to be pre-main-sequence.

A more graphic presentation of the same results is given in Figure 5.
The upper panel compares the mass ratios, as derived both empirically
(vertical axis) and from our tracks (horizontal axis). For both values,
we include the uncertainties listed in Table 1. While the empirical mass
ratios are known quite accurately, we note the relatively large
uncertainties in the theoretical values, especially for P1540 and
V773~Tau. Both systems lie in regions of the HR diagram where the tracks
are especially crowded, so that modest changes in effective temperature
translate into large mass shifts. The overall agreement between
theoretical and empirical ratios is good, but there is a systematic
drift from unity for higher-mass systems. We attribute this trend to the
difficulty in determining the secondary properties when the primary is a
luminous, massive star. On the other hand, we have not considered 
the effect of convective overshooting, which could be an important physical
process to include in more realistic models of higher-mass stars
(e.g., Ribas et al. 2000).

We have also obtained absolute mass values for the sample of stars
recently investigated by Simon et al. (2000). These authors used
interferometric, molecular-line observations of disks around 9 stars
in Taurus-Auriga. Assuming the disks to be in Keplerian motion, they
were able to ascertain the masses of the central objects. We have
placed all these objects in the HR diagram, and find masses that agree
with those of Simon et al. to within 8~percent. Two exceptions, where we 
find substantially lower mass-values, are the single star BP~Tau and the 
binary UZ~Tau. In both cases, the molecular-line observations are 
problematic, as discussed in detail by Simon et al.

An interesting dividend from our study is the determination of
contraction ages for each object above the ZAMS. As seen in the lower
panel of Figure 4, there is a remarkably close match between the 
ages of the primary and secondary star within each binary. We stress
that there is no reason {\it a priori} to expect such agreement, and
there is no generally accepted theory of binary formation that predicts
coevality. Nevertheless, our finding extends the important, earlier
investigations by Hartigan et al. (1994) and Brandner \& Zinnecker (1997),
which demonstrated a similar trend in visual binaries. The implication of
all these studies is that the individual components of binaries are born
{\it in situ} at the same time, regardless of their eventual configuration.
Now the widest pre-main-sequence binaries have separations of order 
$10^4$~AU (e.g., Mathieu 1994), a distance comparable to the sizes of
dense cloud cores (Myers \& Benson 1983). It is an attractive hypothesis,
therefore, that both components form through protostellar collapse within
the same cloud fragment. A question of key importance for theory is how
the structure and evolution of these molecular structures promotes 
simultaneous density buildup in at least two internal locations.

Returning to the stellar mass determinations, the tally of double-lined
spectroscopic binaries is richer than that studied here. We have omitted a
number of systems with mass ratios very close to unity, since they do not
provide stringent tests on the accuracy of our tracks. These binaries include
DQ~Tau ($q=$1.03; Mathieu et al.  1997), W134 ($q=$1.04, Padgett \&
Stapelfeldt 1994), and V826~Tau ($q=$1.02, Lee et al. 1994).  Adopting the
quoted luminosities and temperatures for each star, we concur that the
component masses all match to within 10~percent. We have also omitted the
equal-mass, eclipsing binary GG Ori in the Orion OB1 association. Our mass
estimate of $2.36~\Msun$ for each component is very close to the value of
$2.34~\Msun$ found by Torres et al. (2000). Finally, we have not analyzed 
RXJ 0529.4+0041, an eclipsing binary in the Orion OB1a subgroup
(Covino et al. 2000). The measured masses are $1.30$ and $0.95~\Msun$, but
the luminosity and temperature of the secondary are not yet firmly
established. Our own value for the primary mass is $1.26~\Msun$, in close
agreement with the observational finding.

Our study by no means represents a thorough empirical test of 
pre-main-sequence tracks. Future investigations would do well to
concentrate on lower-mass stars, and on systems with larger mass ratios
than those considered here. As we have discussed, the largest uncertainties
in modeling are for objects near the brown dwarf limit. Traditional
studies of spectroscopic binaries have also emphasized nearly
equal-mass systems, simply because of the practical difficulty in
detecting a relatively dim companion. This purely observational bias has
misled theorists considering the question of binary origin (e.g., Boss
1988, 1993). Recently, Mazeh et al. (2000) have stressed the utility of
near-infrared observations for ascertaining the properties of lower-mass
secondaries. With a wider range of binary systems in hand, we may look 
forward to a more complete assessment of current pre-main-sequence tracks,
and to the eventual refinement of the underlying theory.

\bigskip
\bigskip
\bigskip

\acknowledgements
We are grateful to E.~Covino, S.~Shore, and H.~Zinnecker for very useful
discussions on young binaries.  We also thank the referee, I. Baraffe,
for a very thoughtful report. The funding of F.~P. was partially supplied by
ASI Grant ARS~98-116 to the Osservatorio di Arcetri. Support for S.~S. was
through NSF grant AST-9987266. This work has made extensive use of the
SIMBAD and ADS data bases.

\newpage
 
%
% Figure Captions
%

\figcaption{HR diagram of main-sequence binary systems. Only stars with 
mass smaller than 1.2 $\Msun$ are included. Each evolutionary tracks 
is labeled by the corresponding mass, in solar units.}

\figcaption{The location of the four eclipsing, pre-main sequence systems in
the HR diagram.  In each panel the evolutionary tracks are shown by the solid
lines, labeled by the appropriate mass (in solar units). Dashed lines
are isochrones, given in units of $10^6$ yr. The dotted line is the birthline
computed with an accretion rate $\dot M=10^{-5}$ $\Msun$~yr$^{-1}$.  In the
case of Rs~Cha, the shaded area represents the instability strip for young
stars.}

\figcaption{The position of the non-eclipsing, spectroscopic binaries in the
HR diagram.  The curves have the same meaning as in Fig.~2.  For the GG Tau
quadruple system, the faintest component lies below our minimum mass track
of $0.1\,\Msun$.}

\figcaption{Comparison between theoretical and empirical mass and age
estimates.  The upper panel shows the mass ratios. Here, all binaries are
included except GG Tau, for which mass ratios cannot be obtained
observationally.  The error bars are as indicated in Table 1. The lower panel
displays primary and secondary ages. We have omitted such systems as EK Cep,
whose primary is located on the ZAMS, and thus has no assigned contraction
age.}

\clearpage %
% Tables 
\begin{deluxetable}{lccccccc} \footnotesize
\tablecaption{Parameters of Young Binary Systems}
\tablewidth{0pt}\tablehead{ 
\colhead{System} & \colhead{T$_{\rm eff}$} & \colhead{L} & 
\colhead{M$_{\rm dyn}$} & \colhead{M$_{\rm PS}$} & 
\colhead{(M$_A$/M$_B$)$_{\rm dyn}$} & \colhead{(M$_A$/M$_B$)$_{\rm PS}$} & 
\colhead{t$_{\rm PS}$} \\ 
\colhead{(K)}& \colhead{(L$_\odot$)} & \colhead{(M$_\odot$)} &
\colhead{(M$_\odot$)} & \colhead{(M$_\odot$)} & \colhead{(M$_\odot$)} & 
\colhead{(10$^6$ yr)} }  
\startdata 
{\bf RS~Cha} & $d=97~pc$ & & & & & & \nl
A & 7810$\pm$180 & 15.2$\pm$1.2 & 1.858$\pm$0.016 & 1.88 & 1.02$\pm$0.02 & 
   1.04$\pm$0.06 & 5.0$\pm$0.3 \nl
B & 7295$\pm$170 & 13.9$\pm$1.2& 1.821$\pm$0.018 & 1.80 & & & 
   4.3$^{+0.8}_{-0.6}$ \nl
 & & & & & & & \nl
{\bf EK~Cep} & $d=150~pc$ & & & & & & \nl
A & 9000$\pm$200 & 14.8$\pm$1.4 & 2.03$\pm$0.01 & 1.97 & 1.82$\pm$0.02 & 
   1.73$^{+0.15}_{-0.05}$ & 20\nl
B & 5700$\pm$200 & 1.55$\pm$0.3 & 1.12$\pm$0.01 & 1.14 & & & 20$\pm$4\nl
 & & & & & & & \nl
{\bf TY~CrA} & $d=130~pc$ & & & & &  \nl
A & 12000$\pm$500 & 67$\pm$12 & 3.16$\pm$0.02 & 2.91 & 1.93$\pm$0.02 & 
   1.82$^{+0.11}_{-0.04}$ & -- \nl
B&4900$\pm$400 & 2.4$\pm$0.8& 1.64$\pm$0.01 &1.60& & & 3.9$^{+3.6}_{-2.4}$\nl
 & & & & & & & \nl
{\bf BM~Ori} & $d=460~pc$ & & & & &  \nl
A & 19950$\pm$200 & 1890$\pm$250 & 6.3$\pm$0.3 & 6.21 & 2.52$\pm$0.15 & 
   2.32$\pm$0.15 & $<$0.1 \nl
B & 5740$\pm$200 & 31$\pm$4 & 2.5$\pm$0.1 & 2.68 & & & $<$0.1 \nl
 & & & & & & & \nl
{\bf V773~Tau} & $d=148~pc$ & & & & & & \nl
A & 4900$\pm$180 & 2.31$\pm$0.43 & & 1.53 & 1.32$\pm$0.06     & 
   1.29$^{+0.35}_{-0.23}$ & 4.1$\pm$0.9 \nl
B & 4400$\pm$180 & 1.39$\pm$0.43 & & 1.19 & & & 2.8$^{+1.6}_{-1.0}$ \nl
 & & & & & & & \nl
{\bf NTTS~162814} & $d=125~pc$ & & & & & & \nl
A &4390&1.20& & 1.26 & 1.09$\pm$0.07 & 1.11$\pm$0.07 & 5.1$^{+0.2}_{-2.1}$ \nl
B & 4083 & 0.85 & & 1.08 & & & 5.3$^{+2.5}_{-0.1}$ \nl
 & & & & & & & \nl
{\bf P1540} & $d=470~pc$& & & & & & \nl
A & 5010 & 11.5 & & 1.71 & 1.32$\pm$0.03 & 1.19$\pm$0.18 & $<$0.1 \nl
B & 4680 & 7.9  & & 1.25 &               &      & 0.2$\pm$0.1 \nl
 & & & & & & & \nl
{\bf GG Tau} & $d=140~pc$ & & & & & & \nl
Aa & 4000$\pm$150 & 0.84$\pm$0.13 & 1.28$\pm$0.07 & 0.78 & & & 2.3$\pm$0.3 \nl
Ab & 3760$\pm$120 & 0.71$\pm$0.09 & (total)  & 0.54 & & & 1.5$^{+0.2}_{-0.5}$\nl
Ba & 2985$\pm$180 & 0.08$\pm$0.02 &               & $<$0.1   & & &  \nl
Bb & 2690$\pm$190 & 0.015$\pm$0.004 &             & $\ll$ 0.1& & &  \nl
\enddata 
\end{deluxetable}

\clearpage

\begin{figure}[t]
\centerline{\psfig{figure=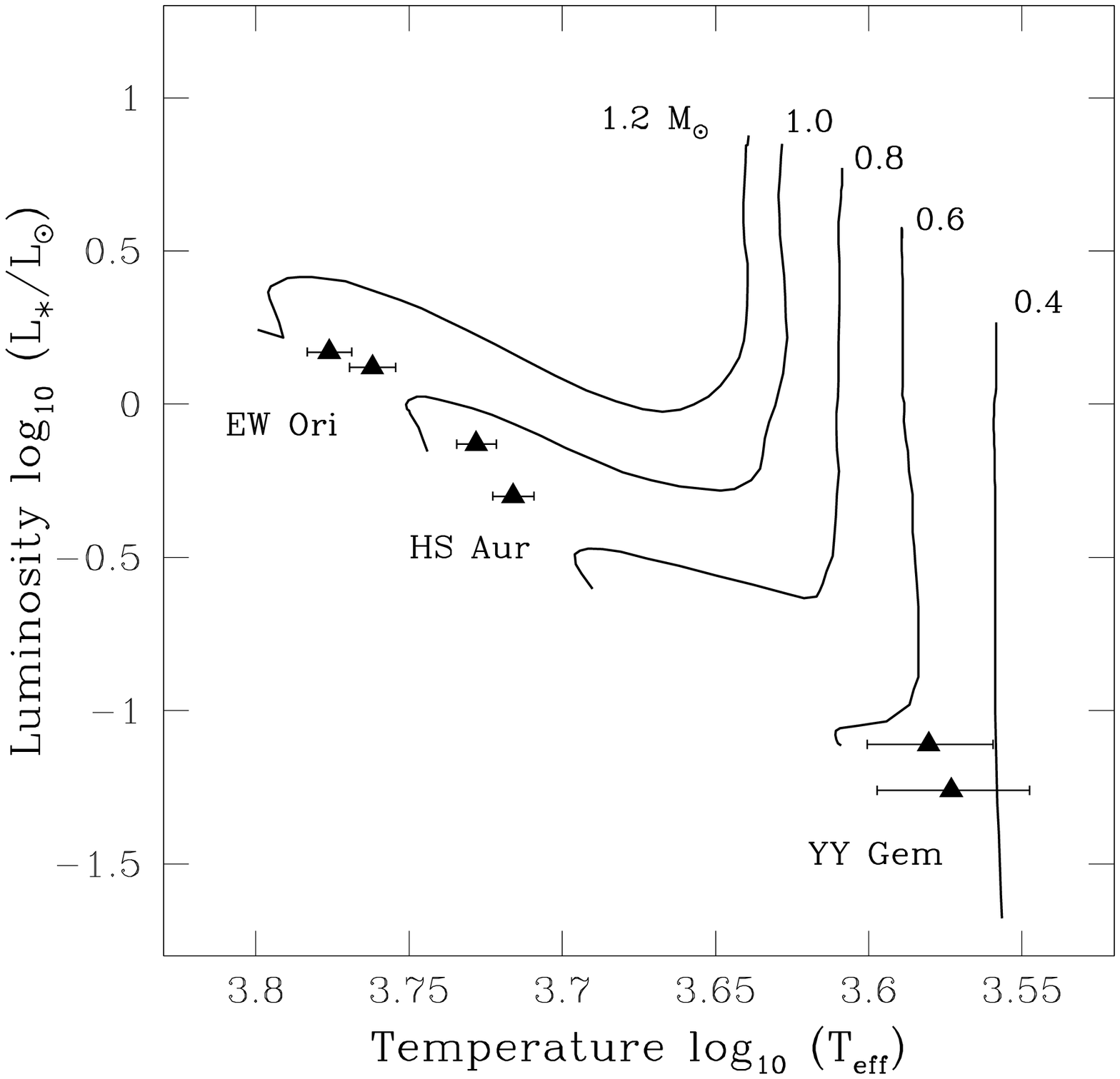}}
\end{figure}

\begin{figure}[t]
\centerline{\psfig{figure=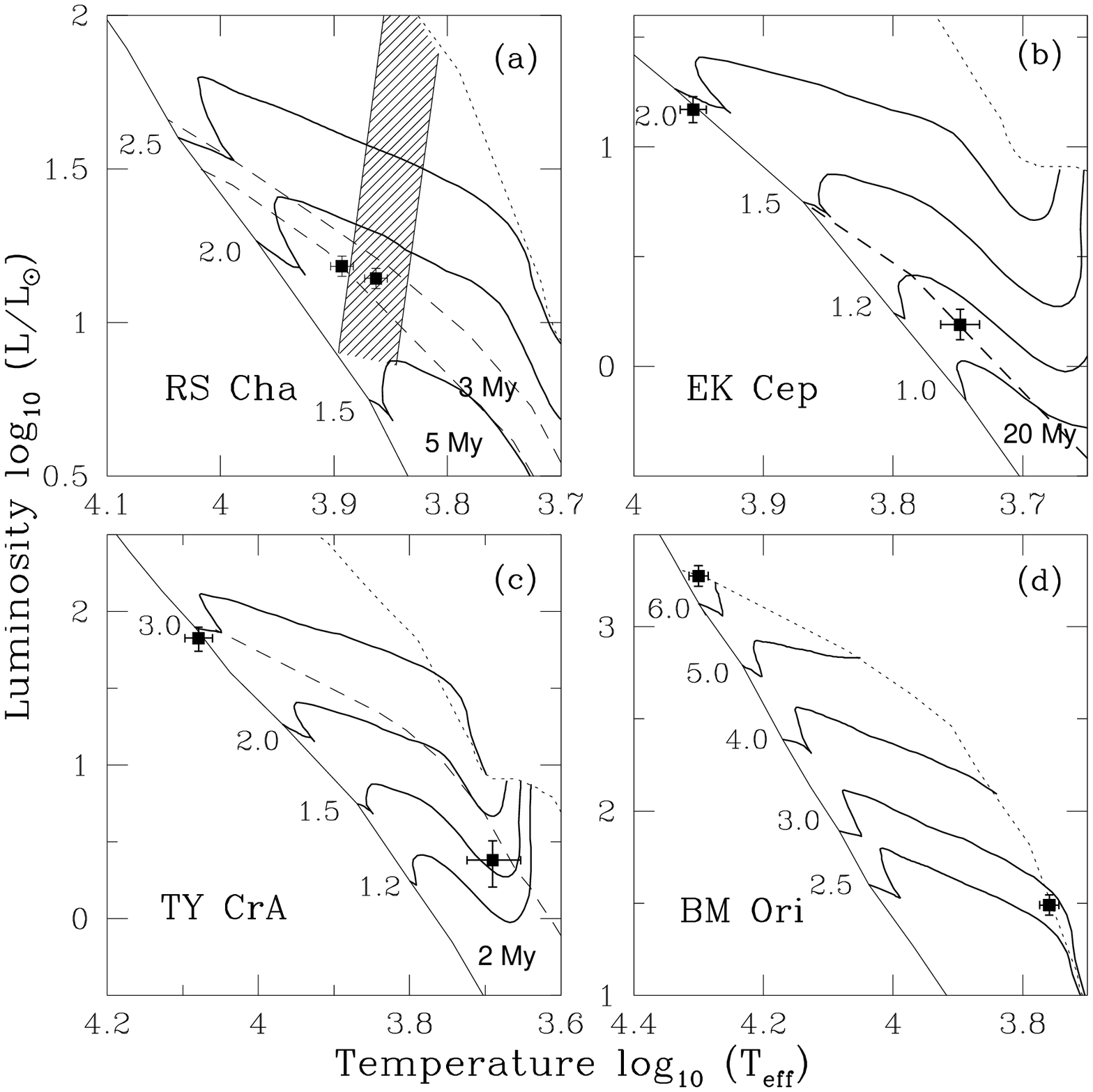}}
\end{figure}

\begin{figure}[t]
\centerline{\psfig{figure=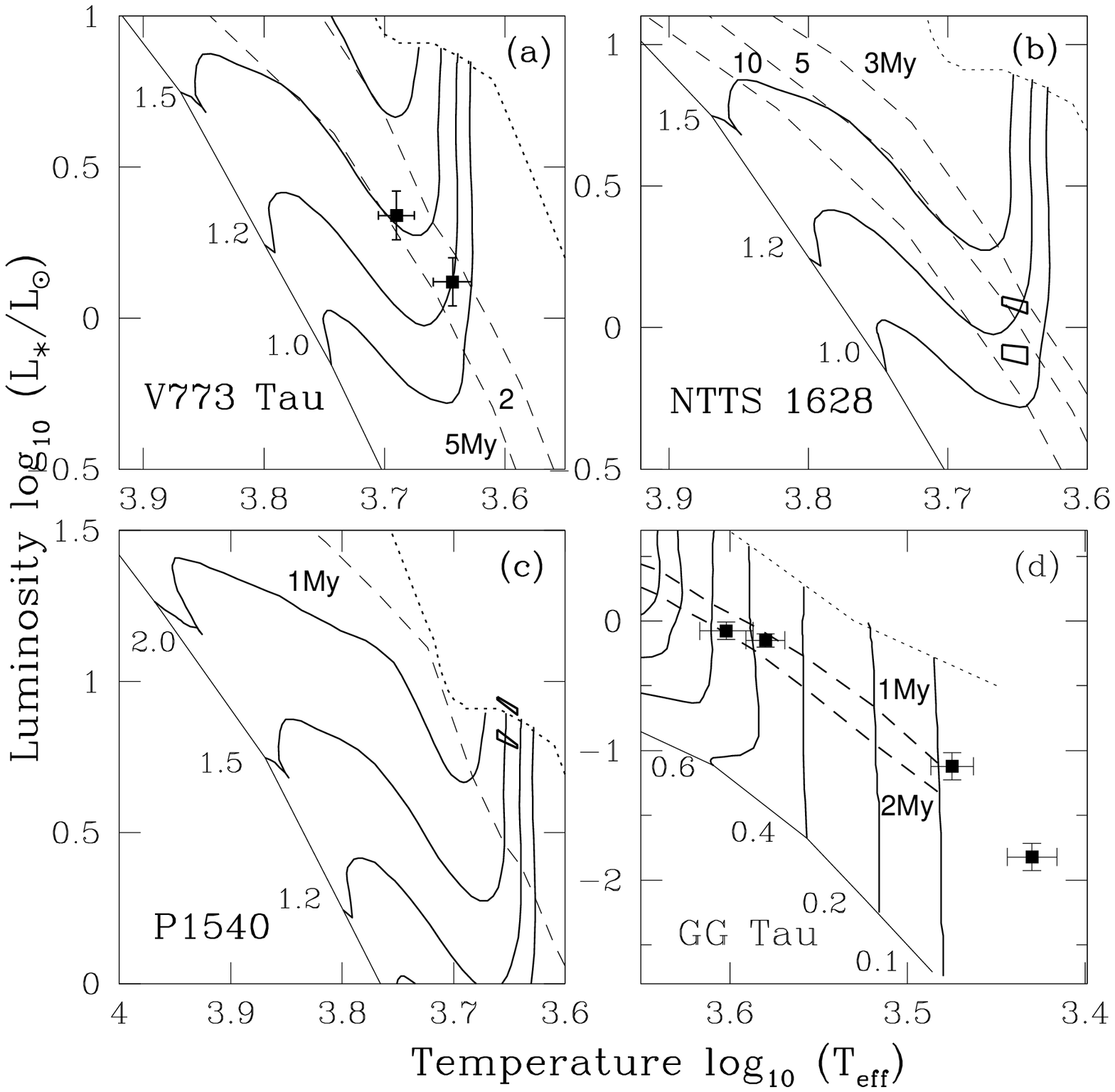}}
\end{figure}

\begin{figure}[t]
\centerline{\psfig{figure=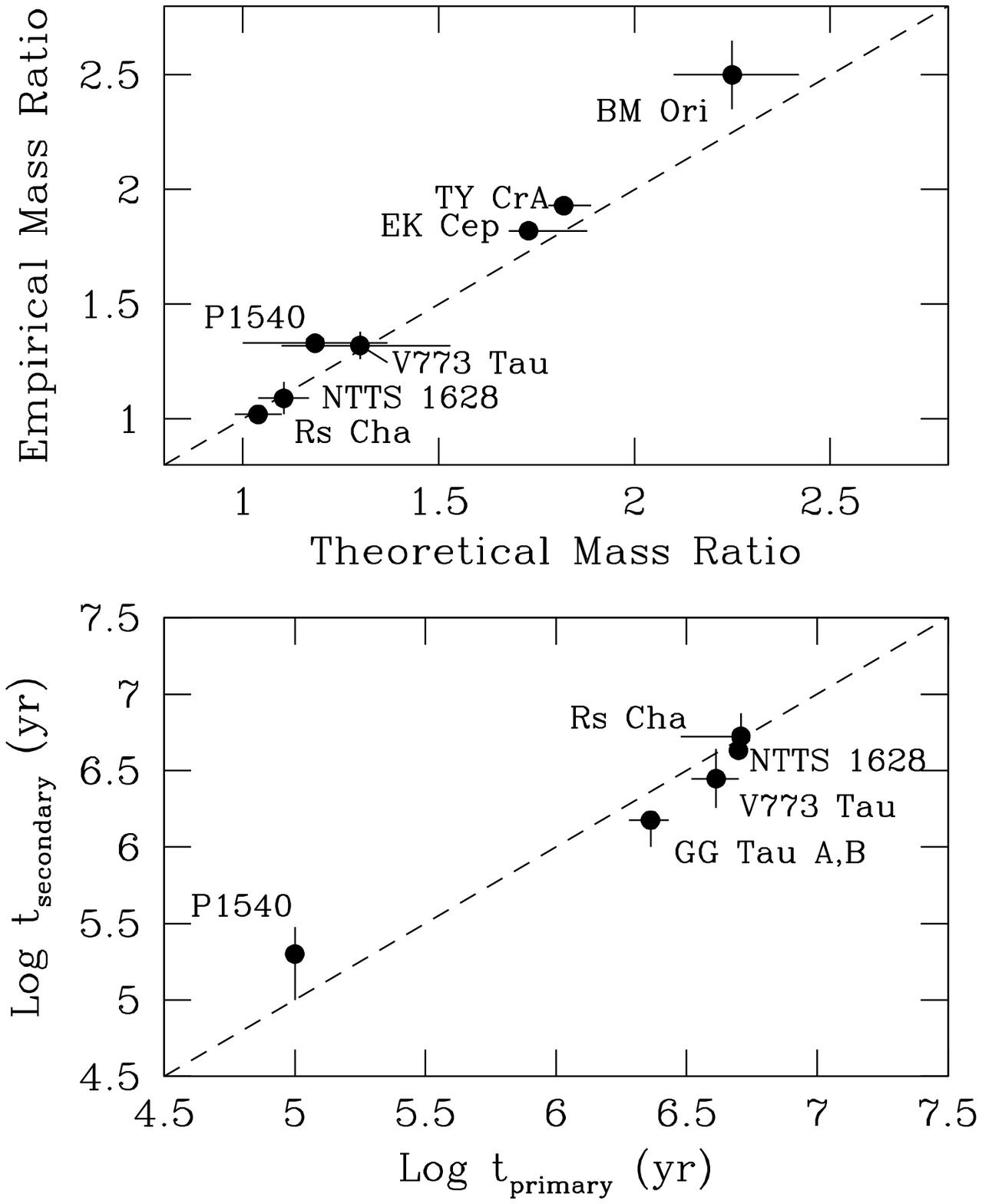}}
\end{figure}

\clearpage

\end{document}